\documentclass[pra, showpacs]{revtex4}
\usepackage{amssymb}
\usepackage{bm}
\usepackage{graphicx}%

\begin{document}
\title{Bose-Einstein condensation in a decorated lattice: an
application to supersolid}

\author{D.\ V.\ Fil$^{1}$ and S.\ I.\ Shevchenko$^2$}

\affiliation{%
$^1$Institute for Single Crystals, National Academy of Sciences of
Ukraine, Lenin av. 60, Kharkov 61001, Ukraine\\ $^2$B.\, Verkin
Institute for Low Temperature Physics and Engineering, National
Academy of Sciences of Ukraine, Lenin av. 47 Kharkov 61103,
Ukraine}

\begin{abstract}
The Bose-Einstein condensation of vacancies in a three-dimensional
decorated lattice is considered. The model describes possible
scenario of superfluidity of solid helium, caused by the presence
of zero-point vacancies in a dislocation network. It is shown that
the temperature of Bose-Einstein condensation decreases under
increase of the length of the segments of the network, and the law
of decrease depends essentially on the properties of the vertexes
of the network. If the vertexes correspond to barriers with a
small transparency, the critical temperature is inversely as the
square of the length of the segment. On the contrary, if the
vertexes correspond to traps for the vacancies (it is
energetically preferable for the vacancies to localize at the
vertexes), an exponential lowering of the temperature of
transition takes place. The highest temperature of Bose-Einstein
condensation is reached in the intermediate case of vertexes with
large transparency, but in the absence of tendency of localization
in them. In the latter case the critical temperature is inversely
as the length of the segment.
\end{abstract}
 \maketitle

\section {Introduction}

Experimental observation of non-classical rotational inertia in
torsion experiments on solid He \cite {1} (confirmed by a number
of other groups \cite {2,3,4}) has revived interest to the idea on
supersolid. The idea goes back to pioneer work by Andreev and
Lifshitz \cite {5} where it was shown that the presence of
vacancies in quantum crystals at zero temperature (zero-point
vacancies) can cause superfluid properties of such systems.
However, as was found later \cite {6-1,7-1}, in $^4$He crystals
the concentration of vacancies is negligible small and distincts
from zero only due to thermal excitation. The presence of
extensive defects  in the crystal can change the situation and
make the occurrence of zero-point vacancies energetically
favorable. If an extensive defect is homogeneous, zero-point
vacancies can move freely along the defect. At sufficient
concentration of defects they form a network, that provides
possibility of flowing of the vacancies through the whole crystal.
Under lowering of the temperature such a gas of vacancies should
go into a superfluid state. In recent papers \cite {13,14} it was
established by Monte Carlo simulation that in $^4$He crystals the
grain boundaries and dislocations do possess superfluid
properties. On the other hand, as was shown in \cite {2},  a
rather long-term annealing (that removes dislocation from the
crystal) leads to a complete disappearance of the effect of a
step-like change of the period of oscillation of a torsion
pendulum filled with solid helium. Thus, superfluidity of
vacancies in a network of dislocations can be considered as
probable  mechanism of superfluidity of quantum crystals.

The idea on dislocation superfluidity was put forward  in Refs.
\cite {15,16} long before the observation of the effect \cite {1}.
As was shown in \cite {15,16}, the important parameter that
determines the temperature of the phase transition in such a
system is the length of the segment of the network. However the
approach \cite {15,16} did not consider such characteristic of the
network as the transparency of the vertexes  (more precisely,
there was some implicit assumption on its value).

In this paper we consider a simple model of the network that
allows to investigate the dependence of the temperature of the
Bose-Einstein condensation on the length of the segment and on
tunnel characteristics of the vertexes.

\section {The model}

\begin {figure}
\begin {center}
  \includegraphics [width=8cm] {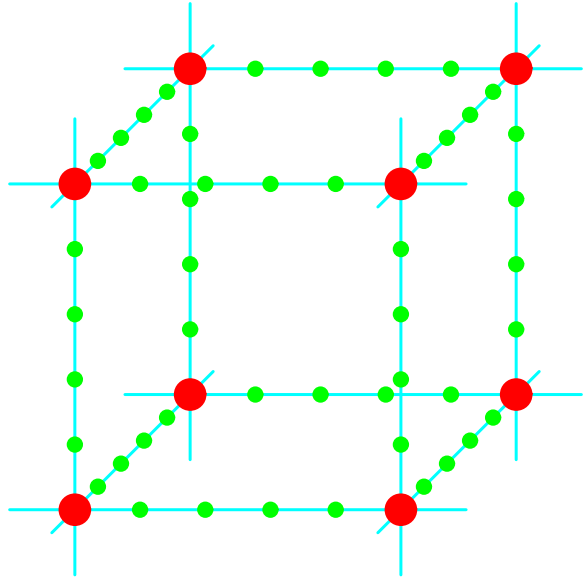}
  \caption {\label {f1} Elementary cell of the decorated lattice ($q=5$).
  Red circles represent central sites,
  green ones, the internal sites of the segments}
  % \label {w1}
\end {center}
\end {figure}

In what follows we will model a superfluid dislocation  as
one-dimensional lattice chain with the period $a$ and the length
$l=q a $. It is implied that the chain has nonzero concentration
of zero-point vacancies $n_1$ (one-dimensional concentration). The
chains (segments) are joined into a regular three-dimensional
network. Two edge sites of each segment are the vertexes of the
network (we will  call such sites the central ones). To be more
specific, we consider that the network  obeys cubic symmetry. In
fact, we model the network as a decorated cubic lattice with the
period $l$ and the number of sites in the elementary cell equal
$3q-2$ (Fig. \ref{f1}). We are interested in the case of large
$q$. The vacancies that moves in a such lattice are described in
the tight-binding approximation. The model contains two
parameters: $t$, the amplitude of tunnelling between nearest
neighbor internal sites in segments, and $t_1$, the amplitude of
tunnelling between a central site and a nearest neighbor internal
site. The one-site energies are assumed to be the same for all
sites. In the broader sense, the model describes a network formed
by one-dimensional wires, along which bosons can move. The central
sites play the role of scatterers that connect the wires.

The number of zero-point vacancies $N$ is supposed to be much
smaller than the numbers of sites in the decorated lattice, i.e.
the filling factor for the vacancies satisfies the inequality $
\nu=n_1 a\ll 1$. For the further analysis it is convenient to
introduce three-dimensional density of vacancies $n\approx 3
q\nu/l^3$. We note that even at small filling factor $\nu\ll 1$
the number of vacancies per elementary cell can be much larger
than unity ($n l^3\gg 1$) if the condition $\nu\gg 1/q $ is
satisfied. Below we will consider the filling factors  belonging
to the diapason $1/q\ll \nu\ll 1$.

At small filling factors one can neglect the interaction between
the vacancies. The temperature of Bose-Einstein condensation $T_0$
for a non-interacting gas of bosons in a compound lattice is
determined by the equation
\begin {equation} \label {1}
    N =\sum_{\lambda, {\bf k}}
    \frac {1} {\exp\left (\frac {\epsilon_\lambda (\bf k)-\epsilon_0} {T_0} \right)-1},
\end {equation}
where $ \epsilon_\lambda (\bf k) $ is the spectrum of bosons in a
lattice, $ \lambda $, the band index, $ {\bf k} $, the wave
vector, $ \epsilon_0$, the energy that corresponds to a bottom of
the lowest band. In (\ref {1}) the index $\lambda$ runs from 1 to
$3q-2$ (the number of sites in the unit cell), and summation in
taken over $ {\bf k} $ belonging to the first Brillouin zone
($-\pi/l <k_i <\pi/l $).

The Hamiltonian of the system has the form
\begin {eqnarray} \label {2}
    H = - \sum _ {\bf i} \sum _ {\alpha=x, y, z} [t_1 (b _ {{\bf i}, v} ^ +
    b _ {{\bf i}, (\alpha, 1)} + b _ {{\bf i}, (\alpha, q-1)} ^ +
    b _ {{\bf i} + {\bf n} _ \alpha, v}) +
     t\sum _ {\xi=1} ^ {q-2} b _ {{\bf i}, (\alpha, \xi)} ^ +
    b _ {{\bf i}, (\alpha, \xi+1)} + h.c.],
\end {eqnarray}
where $b^+_{{\bf i}, \eta}$ ($b_{{\bf i}, \eta} $) is the operator
of creation (annihilation) of a boson in the site $\eta$ in the
$i$-th cell, $ {\bf i} $, the radius-vector of the $i$-th cell, $
{\bf n} _x = (l, 0,0) $, $ {\bf n} _y = (0, l, 0) $, $ {\bf n} _z
= (0,0, l) $, the primitive vectors of translation. The following
notation for $ \eta $ is used: $ \eta=v $, the central site, $
\eta = (\alpha, \xi) $, the $ \xi $-th internal internal site in a
segment aligned in $\alpha $ direction.

Applying the Fourier-transformation
\begin {equation} \label {5}
    b _ {{\bf i}, \eta} = \frac {1} {\sqrt {N_i}} \sum _ {\bf k} b _ {{\bf k}, \eta} e ^ {i {\bf k i}}
\end {equation}
(where $N_i $ - number of unit cells), we rewrite the Hamiltonian
as
\begin {equation} \label {6}
    H =\sum _ {\eta_1, \eta_2} M _ {\eta_1, \eta_2} ({\bf k}) b _ {{\bf
    k}, \eta_1} ^ + b _ {{\bf k}, \eta_2}.
\end {equation}
The matrix $ \mathbf {M} ({\bf k}) $ has dimension $ (3q-2) \times
(3q-2) $ and is presented in the following block form:
\begin{equation}\label{7}
    \mathbf{M}({\bf k})=-t\left(%
\begin{array}{cccc}
  0 & \mathbf{T}_x & \mathbf{T}_y  & \mathbf{T}_z  \\
  \mathbf{T}_x^+  & \mathbf{D}_{q-1} & \mathbf{0} & \mathbf{0} \\
   \mathbf{T}_y^+ & \mathbf{0}  & \mathbf{D}_{q-1}  & \mathbf{0}  \\
   \mathbf{T}_z^+  & \mathbf{0}  & \mathbf{0}  & \mathbf{D}_{q-1} \\
\end{array}%
\right)
\end{equation}
Here $ \mathbf {D} _ {q-1} $ is the $ (q-1) \times (q-1) $ matrix
that corresponds to the tunnelling between internal sites:
\begin{equation}\label{8}
\mathbf{D}_{q-1}=\left(%
\begin{array}{ccccc}
  0 & 1 & \ldots & 0 & 0 \\
  1 & 0 & \ldots & 0 & 0 \\
  \ldots & \ldots & \ldots & \ldots & \ldots \\
  0 & 0 & \ldots & 0 & 1 \\
  0 & 0 & \ldots & 1 & 0 \\
\end{array}%
\right)
\end{equation}
and $ \mathbf {T} _ \alpha $ and $ \mathbf {T} _ \alpha ^ + $ are
of 1$ \times (q-1) $ and $ (q-1) \times 1$ matrixes that describe
the tunnelling between the central site and the nearest internal
site:
\begin{equation}\label{9}
\mathbf{T}_\alpha=\left(%
\begin{array}{ccccc}
  \tau & 0 & \dots & 0 & \tau e^{-i k_\alpha l} \\
\end{array}%
\right), \qquad \mathbf{T}_\alpha^+=\left(%
\begin{array}{c}
  \tau \\
  0 \\
  \ldots \\
  0 \\
  \tau e^{i k_\alpha l} \\
\end{array}%
\right)
\end{equation}
$ \tau=t_1/t $ is a key parameter of the model (the ratio of
amplitudes of tunnelling
 between an internal site and a central site and between two internal sites).

The spectrum of bosons satisfies the dispersion equation $ \det
(\epsilon \mathbf {I} - \mathbf {M}) =0$. Using the explicit
expressions for $ \mathbf {M}$ (\ref{7})-(\ref{9}), we obtain the
explicit form of the dispersion equation
\begin {eqnarray} \label {10}
    [\Delta _ {q-1} (\tilde {\epsilon})] ^2 [\tilde {\epsilon} \Delta _ {q-1} (\tilde {\epsilon})-6
    \tau^2 \Delta _ {q-2} (\tilde {\epsilon}) - (-1) ^q 2 \tau^2 \sum_\alpha
\cos (k_\alpha l)] =0,
\end {eqnarray}
where $ \tilde {\epsilon} = \epsilon/t $ and $ \Delta _ {q}
(\tilde {\epsilon}) = \det (\tilde {\epsilon} \mathbf {I} +
\mathbf {D} _ {q}) $.

\section {Band structure of the spectrum}

As follows from the dispersion equation (\ref {10}), some bands
are reduced to degenerate levels. The energies of these levels are
given by the equation
\begin {equation} \label {10-1}
\Delta _ {q-1} (\tilde {\epsilon}) =0.
\end {equation}
Eq. (\ref{10-1}) coincides with the dispersion equation for an
isolated chain with  $q-1$ sites. The energies  of the levels are
equal to
\begin {equation} \label {14}
    \epsilon_s =-2 t \cos\frac {\pi s} {q}
\end {equation}
($s=1,2, \ldots, q-1$). The degree of degeneracy of each level is
$2N_i$. The wave functions of such states has  zero  weight in the
central sites that is in agreement with the absence of dispersion.

\begin {figure}
\begin {center}
  \includegraphics [width=14cm] {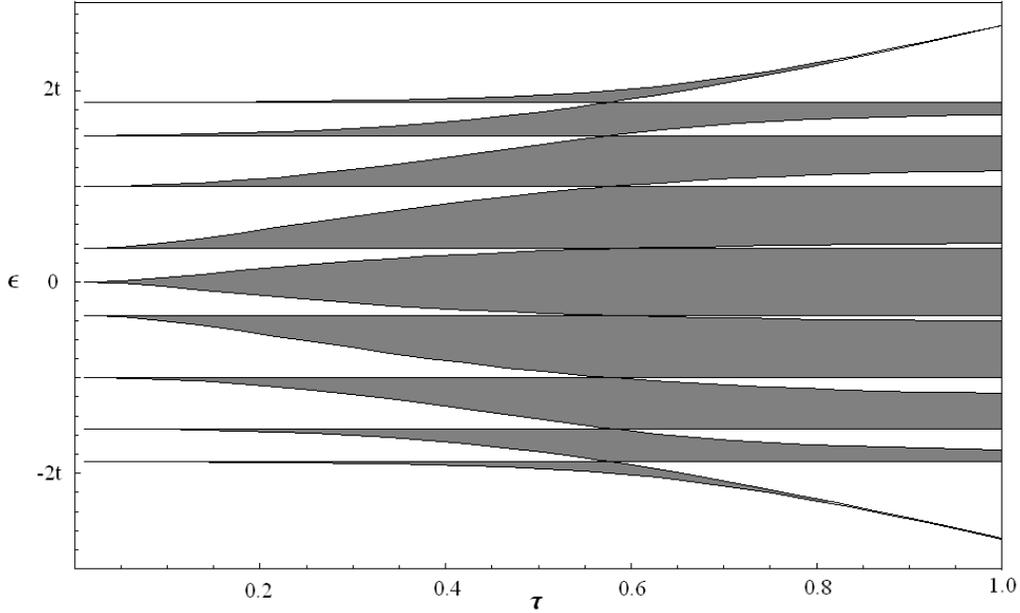}
  \caption {\label {f2} The band structure for bosons in the decorated lattice (for $q=9$).
  The allowed bands are shown grey.}
  % \label {w1}
\end {center}
\end {figure}

The spectrum of the bands with a finite dispersion satisfies the
equation
\begin {equation} \label {14-1}
\tilde {\epsilon} \Delta _ {q-1} (\tilde {\epsilon})-6
    \tau^2 \Delta _ {q-2} (\tilde {\epsilon}) - (-1) ^q 2 \tau^2 \sum_\alpha
\cos (k_\alpha l) =0.
\end {equation}
 The left hand part of Eq. (\ref {14-1}) is a $q$-th order polynomial.
 It has $q$ distinct solutions that corresponds to $q$ bands (allowed bands).
The widths of the bands essentially depend on the parameter
$\tau$.  It is illustrated in  Fig. \ref{f2}.

One can see from this figure, that there is a special value $\tau
=\tau_c=1/\sqrt {3}$ at which all allowed bands have the largest
width, and there is no gaps between the bands. At $\tau>\tau_c$
and $\tau <\tau_c$ the allowed bands becomes narrower and energy
gaps (forbidden bands) appear. The effect of narrowing  is
strongest for the lowest and the highest bands. As we will see
later, the temperature of Bose-Einstein condensation is determined
in the main part  by the width of the lowest band. On a
qualitative level the dependence of the width of bands on $\tau$
is explained as follows. At small $ \tau $ it is energetically
preferable for the vacancies to  localize inside segments, and the
central sites play the role of the barriers with a small
transparency. As is known, in particular, from the Kronig-Penny
model (see, for example, \cite {17}), in such a situation a sharp
narrowing of the lowest band takes place. At $\tau> \tau_c $ it is
energetically preferable for the vacancies to localize at central
sites, and overlapping between such localized states becomes very
small.

To obtain quantitative estimation for the Bose-Einstein
condensation temperatures it is necessary to find the spectrum of
the bands for an arbitrary $q$. From the definition of $ \Delta_q
(\tilde {\epsilon}) $ follows that this function satisfies the
recurrent relation
\begin {equation} \label {11}
\Delta_q (\tilde {\epsilon}) = \tilde {\epsilon} \Delta _ {q-1}
(\tilde {\epsilon}) - \Delta _ {q-2} (\tilde {\epsilon})
\end {equation}
($ \Delta_1 (\tilde {\epsilon}) = \tilde {\epsilon} $, $ \Delta_2
(\tilde {\epsilon}) = \tilde {\epsilon} ^2-1$). Using the relation
(\ref {11}) and applying the method of a mathematical induction
one can prove  that
\begin {equation} \label {12}
    \Delta_q (2 \cos \gamma) = \frac {\sin [(q+1) \gamma]} {\sin\gamma}
\end {equation}
and
\begin {equation} \label {13}
    \Delta_q (\pm 2  \cosh \gamma) =
    (\pm 1) ^q \frac {\sinh [(q+1) \gamma]} {\sinh \gamma}.
\end {equation}
Eqs. (\ref {12}) and (\ref {13}) allow to reduce the equations
(\ref {10-1}) and (\ref {14-1}) to compact trigonometrical
equations. In particular, the substitution $ \tilde {\epsilon} =2
\cos \gamma $ reduces Eq. (\ref {10-1})  to  the equation $ \sin q
\gamma =0$ (with the additional condition $ \sin\gamma \ne 0$)
which solutions correspond to the energies (\ref {14}).

For finding the spectrum  we use the substitution $ \tilde
{\epsilon} =2 \cos (\pi +\gamma) $ that reduces Eq. (\ref {14-1})
to the following equation for $ \gamma $:
\begin {equation} \label {212}
    \cot \gamma \sin (q\gamma) (1-3\tau^2) +3\tau^2
    \left (\cos (q\gamma)-\frac {1} {3} \sum_\alpha
\cos (k_\alpha l) \right) =0.
\end {equation}
At $ \tau =\tau_c=1/\sqrt {3} $ Eq. (\ref{212}) can be easily
solves that yields the spectrum
\begin {equation} \label {23}
    \epsilon _ {j} ({\bf k}) =-2t\cos
    \left (\frac {2\pi} {q} \left [\frac {j} {2} \right] - (-1) ^j\frac {1} {q} \arccos\frac {\sum_\alpha
\cos (k_\alpha l)} {3} \right)
\end {equation}
($j=1,2, \ldots q $), where square brackets indicate the integer
part. One can see from (\ref {23}) that at any $q$ the bandgaps
are equal to zero. We display also approximate expression for the
half-width of lower bands
\begin {equation} \label {23-1}
    W_j\approx \frac {t \pi^2} {q^2} (j-\frac {1} {2})
\end {equation}
($j\ll q $). As follows from (\ref {23-1}), the half-widths are in
inverse proportion with the square of length of the segment (at
fixed $a$).

At $ \tau <\tau_c $ one can find an approximate solutions of Eq.
(\ref {212}) in a diapason of small $ \gamma $ (that correspond to
the lower bands). We specify the case of $ \tau $ not too close to
$ \tau_c $, when the condition $ (1-3\tau^2) q\gg 1$ is satisfied.
The solution of Eq. (\ref {212}) can sought as $ \gamma =\gamma _
{j} + \tilde {\gamma} _j $, where $ \gamma_j =\pi j/q $ ($j\ll q
$) and $ \tilde {\gamma} _j\ll \gamma _ {j} $. The approximate
expression for $ \tilde {\gamma} _j $ as obtained from (\ref
{212}) reads as
\begin {equation}
\label {213}
 \tilde {\gamma} _j\approx-\frac {3\tau^2} {1-3\tau^2} \frac {\pi
 j} {q^2} \left (1-\frac {(-1) ^ {j}} {3} \sum_\alpha
\cos (k_\alpha l) \right).
\end {equation}
It gives the following expression for the spectrum of lower bands
\begin {equation} \label {26}
    \epsilon _ {j} ({\bf k}) \approx-2t\left [\cos\frac {\pi j} {q}
    + \frac {\pi^2 j^2} {q^3} \frac {3\tau^2} {1-3 \tau^2} \left (1 \frac {(-1) ^ {j}} {3} \sum_\alpha
\cos (k_\alpha l) \right) \right].
\end {equation}
In this case the half-width of the lowest band  is equal to
\begin {equation} \label {214}
    W _ {1} \approx\frac {2 t \pi^2} {q^3} \frac {3\tau^2} {1-3 \tau^2},
\end {equation}
i.e., it is inversely proportional the third power of the length
of the segment. We  note that in the Kronig-Penny model at small
transparency of the barriers the width of the bottom band is also
inversely proportional to the cube of distance between the
barriers (see, for instance, \cite {17}).

If $ \tau $ exceeds $ \tau_c $, (and $q\gg 1$)  a sharp narrowing
of the lowest band takes place, and this band drops below the
level $-2t $. Using the substitution $ \tilde {\epsilon} =-2\
\cosh \gamma $, we obtain from (\ref {14-1}) the following
equation
\begin{equation} \label {206}
\coth \gamma \ \sinh (q\gamma) (1-3\tau^2) +3\tau^2 \left [\cosh
(q\gamma)-\frac {1} {3} \sum_\alpha \cos (k_\alpha l) \right] =0.
\end{equation}
At $q (3\tau^2-1) \gg 1$ it is convenient to seek for a solution
of Eq. (\ref {206}) in the form $ \gamma =\gamma_0 +\tilde
{\gamma} $, where $ \gamma_0$ is given by the equation
\begin {equation} \label {16}
    \coth \gamma_0 = \frac {3\tau^2} {3 \tau^2-1}
\end {equation}
(i.e. $ \gamma_0 =\ln (6\tau^2-1)/2$), and an exponentially small
correction $ \tilde {\gamma}$ can be obtained directly from (\ref
{206})
\begin {equation} \label {208}
\tilde {\gamma} \approx \frac {2} {3} \sinh \gamma_0 \ \cosh
\gamma_0 \ e ^ {-q\gamma_0} \sum_\alpha \cos (k_\alpha l).
\end {equation}
The spectrum of the lowest band reads as
\begin {equation} \label {19}
    \epsilon _ {1} ({\bf k}) \approx-E_0
    \left (1 +\frac {2\sinh ^2 \gamma_0} {3} e ^ {-q\gamma_0} \sum_\alpha
\cos (k_\alpha l) \right),
\end {equation}
where $$ E_0=2t \cosh \gamma_0=2t\frac {3 \tau^2} {\sqrt {6
\tau^2-1}}. $$  Thus, the lowest band becomes exponentially narrow
(with the half-width $W_1\approx 4t \cosh \gamma_0 \sinh ^2
\gamma_0 e ^ {-q\gamma_0} $). All the others bands lay above the
level $-2t $. The spectra for the lower bands (starting from
$j=2$) are determined by the equation (\ref {26}) if one replaces
$j $ for $j+1$ in its left hand part. The eigenfunctions for the
lowest band have the maximum weight in the central site and it
falls quickly with the distance from the central site.

\section {Temperature of Bose-Einstein condensation}

Due to the presence of degenerate levels in the spectrum the
formula (\ref {1}) for the temperature of Bose-Einstein
condensation(BEC) is modified to
\begin {equation} \label {205}
    N =\sum _ {j=1} ^ {q} \sum _ {{\bf k}}
    \frac {1} {\exp\left (\frac {\epsilon_j (\bf
    k)-\epsilon_0} {T_0} \right)-1}
    +2N_i\sum _ {s=1} ^ {q-1} \frac {1} {\exp\left (\frac {\epsilon_s-\epsilon_0} {T_0} \right)-1}.
\end {equation}
We are interested in systems with a rather large concentration of
vacancies ($n\gg l ^ {-3} $).  In this case the BEC temperature is
much larger than the width of the lowest band.  It allows to use
the approximate expression for the Bose distribution function for
such a band
$$ \frac {1} {\exp\left (\frac {\epsilon_1 (\bf
    k)-\epsilon_0} {T_0} \right)-1} \approx \frac {T_0} {\epsilon_1 (\bf
    k)-\epsilon_0}. $$

The rest bands and degenerate levels give non-negligible
contribution into (\ref{205}), only if their energies, counted
from a bottom of the lowest band, are of order or less than the
BEC temperature. The main contribution into (\ref {205}) yields
the lowest band. Therefore, for evaluation of what bands and
levels should be taken into account, one can use the estimate
$T_0\sim W_1 nl^3$.

Let us first consider $ \tau <\tau_c $ (and $q\gg 1 /
(1-\tau^2/\tau_c^2) $). In this range of the parameters the
half-width of the lowest band $W_1\propto q ^ {-3} $, while the
energy gap between the first and second bands $ \Delta _ {g, 1}
\propto q ^ {-2} $. At  small filling factors $\nu $ the
three-dimensional density satisfies the condition $nl^3=3q\nu\ll q
$. Hence, $T_0\ll \Delta _ {g, 1} $ and  it is enough to take into
account in Eq. (\ref {205}) only the lowest band and the lowest
degenerate level (that lies at the top of the lowest band). As a
result, Eq. (\ref {205}) is reduced to
\begin {equation} \label {2006}
    nl^3\approx \frac {1} {\pi^3} \int _ {0} ^ \pi\int _ {0} ^ \pi\int _ {0} ^ \pi
 d \tilde {k} _x d \tilde {k} _y d \tilde {k} _z \frac {T_0} {W_1 [1-\frac {1} {3} (\cos \tilde {k} _x +
 \cos \tilde {k} _y + \cos
\tilde {k} _z)]} + \frac {T_0} {W_1} \approx 2.5 \frac {T_0}
{W_1}.
\end {equation}
It is instructive to write down the answer for the BEC temperature
in terms of one-dimensional concentration of vacancies $n_1$, the
distance between nearest sites $a$, and the length of the segment
$l$:
\begin {equation} \label {1001}
    T_0=1.2 \pi^2 \frac {3 \tau^2} {1-3\tau^2} 2ta^2 \frac {n_1 a} {l^2}.
\end {equation}

At $ \tau\approx\tau_c $ the energy gaps between the bands
approach  zero and one should take into account many  bands and
levels. Since the main contribution into (\ref {205}) is given by
the levels and the band, which energy counted from the bottom of
the lowest band is  smaller than $T_0$, it is enough to take into
account only such levels and bands. Besides, it is possible to
approximate all bands, except the first one, by the degenerate
levels (with degree of degeneracy $N_i$), located between
degenerate levels (\ref{14}). As a result, Eq. (\ref {205}) is
reduced to
\begin {equation} \label {218}
    nl^3\approx \frac {T_0} {W_1} \frac {1} {\pi^3} \int _ {0} ^ \pi\int _ {0} ^ \pi\int _ {0} ^ \pi
 d \tilde {k} _x d \tilde {k} _y d \tilde {k} _z \frac {1} {2\left (\frac {1} {\pi} \arccos\frac {\sum_\alpha
\cos \tilde {k} _ \alpha} {3} \right) ^2}
  +2\sum _ {s=1} ^ {s _ {max}} \frac {T_0} {2W_1 s^2}
    + \sum _ {j=1} ^ {j _ {max}} \frac {T_0} {2W_1 {(j +\frac {1} {2})} ^2},
    \end {equation}
where $s _ {max} $ and $j _ {max} $ - numbers of levels and bands,
which energy is of order of $T_0$. In view of fast convergence of
the sums in (\ref {218}) they can be extended to infinity. The
integral in Eq. (\ref {218}) can be evaluated numerically. As a
result, we obtain $T_0\approx 0.2 W_1 nl^3$.  Using (\ref {23-1})
we find
\begin {equation} \label {220}
    T_0\approx 3 t a^2 \frac {n_1} {l}
\end {equation}

At  $ \tau> \tau_c $ the energy gap that separates the lowest band
from the lowest level and the second band is exponentially large
in comparison with the width of the bottom band. Hence, it is
enough to consider in (\ref {205}) only the lowest band. It gives
$T_0\approx 0.7 W_1 n l^3$. At large $q=l/a $ this temperature is
exponentially small. For example, at $ \tau=1$ the BEC temperature
is equal to
\begin {equation} \label {211}
    T_0 \approx 9 t n_1 l e ^ {-\frac {l \ln 5} {2a}}
\end {equation}
We should note that since the states in the lowest band correspond
to the vacancies localized near the central site, the result (\ref
{211}) is more sensitive to the interaction than results for other
$ \tau $. Due to the tendency to localization at central sites,
the interaction can be neglected, only if number of vacancies per
elementary cell is less or of order of unity. Therefore, the
result (\ref {211}) should be considered as qualitative one.

\section {Conclusion}

The model of decorated lattice considered in this paper describes
three physically distinct situations, depending on value of $
\tau$. The case of small $ \tau $  corresponds the situation, when
intersections of dislocations play the role of barriers with a
small transparency for the vacancies. In this case the BEC
temperature is inversely proportional to the square length of the
segment. The case $\tau> \tau_c $ describes the situation, where
the vacancies tend to localize at the dislocation intersections.
In such a situation the BEC temperature decreases exponentially
with the increase  of the length of the segment. At last, the case
$ \tau\approx\tau_c $ corresponds the situation where vacancies
can move freely through the intersections, and do not localize on
them. Last situation is the most preferable for the BEC: only
linear decrease of the temperature of transition with the increase
of the length of the segment takes place. We consider that the
functional dependences of the BEC temperature on the length of the
segment is more or less universal and does not depend on the
mechanism that causes the appearance of the barriers (or the
centers of localization) on the dislocation intersections.

It is of interest to estimate BEC temperature for the most
preferable case (\ref {220}). If one defines the effective mass of
vacancies on a simple (not decorated) cubic lattice $M ^ *
=\hbar^2/2ta^2$ the result (\ref {220}) can be presented in the
form
\begin {equation} \label {1003}
 T_0\approx \frac {3} {2} \frac {\hbar^2} {M ^ *} \frac {n_1} {l}
\end {equation}
Note that the answer (\ref {1003}) does not contain any parameter
of the decorated lattice. Thus this answer can be applied  to a
more general case, where dislocations can be modelled as quasi
one-dimensional wires in which vacancies can move. The answer
(\ref {1003}) up to the numerical factor of order of unity
coincides with the estimate given in Ref. \cite {16}. Implying
that the effective mass of vacancies is approximately equal to the
mass of  $^4$He atom, we evaluate (\ref {1003}) as $T_0\approx
(n_1/l) \cdot 18$ K $\cdot\AA^2$. For example, at density of
vacancies $n_1=1$ $ \AA ^ {-1} $ (a value obtained in \cite {14})
the temperature $T_0\approx 0.1$ K is reached for the length of
the segment $l=180$ $ \AA $, that corresponds to two-dimensional
density of dislocations $n_d=3\cdot 10 ^ {11} $ cm$^{-2} $.

In conclusion, we  discuss shortly the question on the relation
between the Bose-Einstein condensation and the superfluidity of
vacancies in the network of dislocations. At temperatures lower
than the BEC temperature the long-range phase correlations are
established in the system. In our case the phases on different
segments become correlated. It allows to describe the system
(below $T_0$) by a complex order parameter that reflects the
possibility of non-dissipative flow along the dislocations. Above
$T_0$ the flowing without relaxation is impossible. But the
specifics of the system studied consists in that the temperature
$T_0$ and the temperature of degeneracy for the gas of vacancies
on a segment $T_d $ are quite different: $T_d=t n_1^2 a^2\gg T_0$.
Therefore, the relaxation time for the flow at $T_d\gg T> T_0$;
can be very large, and above $T_0$ a peculiar quasi-superfluid
phase can be realized (see \cite {16}, and also discussion in
\cite {14}). Let us note that in the experiment \cite {18} the
attempt of direct observation of Bose-Einstein condensation in
solid helium was made (through the measurements of temperature
dependence of the kinetic energy per atom). According to the
results of Ref. \cite {18}, in the temperature diapason where
nonclassical rotational inertia is observed (measurements were
done down to 0.07 K), the Bose-Einstein condensate does not arise
yet.

%Рассчитанная in the given work critical temperature
%отвечает to system transition in a condition at which there is a correlation of phases of the vacancies which are
%на various segments. Therefore such transition should be accompanied by occurrence
%сверхтекучих properties at a crystal as whole, and superfluidity can have stationary character.
%Проявление non-stationary superfluid properties
% (as response to variable external influence) it is possible and at more heats
% (see discussion in \cite {14}, and also in \cite {15,16}). Non-stationary effects can be connected
%с local superfluidity in a segment.

This study was supported in part by the CRDF grant No 2853.

%\eject

\end {document}